\documentclass[aps,prl,twocolumn,superscriptaddress,floatfix,nofootinbib,longbibliography,nobalancelastpage]{revtex4-2}

\usepackage{amsmath}
\usepackage{times}
\usepackage{amssymb}
\usepackage{graphicx}
\usepackage{soul}
\usepackage[colorlinks=true,linkcolor=blue, citecolor=blue, urlcolor=blue, bookmarks]{hyperref}

\usepackage[draft]{changes}

\makeatletter
\usepackage[T1]{fontenc}
\usepackage{amsmath,amsfonts,amssymb,amsthm,bbm,bm, braket}
\usepackage{wasysym}
\usepackage{comment}
\usepackage{scalerel}
\usepackage{enumerate}
\usepackage{amssymb}
\usepackage{physics}
\usepackage{graphicx}
\usepackage{bbm}
\usepackage{mathtools}
\usepackage{ifthen}
\usepackage{tensor}
\usepackage{tikz}
\usepackage{tikz-network}
\usepackage{booktabs}
\usepackage{multirow}
\usetikzlibrary{patterns,decorations.pathreplacing}

\usepackage{color}
\usepackage{longtable}
\usepackage[normalem]{ulem} 

\theoremstyle{break}       
\usepackage{color}
\definecolor{myred}{RGB}{232,102,102}
\definecolor{myblue}{RGB}{187,187,255}
\definecolor{myorange0}{RGB}{252,226,5}
\definecolor{myorange0c}{RGB}{255,255,255}
\definecolor{myorange}{RGB}{255,165,0}
\definecolor{mygrey}{RGB}{105,105,105}
\definecolor{OliveGreen}{RGB}{85,107,47}
\definecolor{NavyBlue}{RGB}{50,50,168}
\definecolor{mygreen}{RGB}{34,139,34}
\definecolor{myY}{RGB}{220,255,203}
\definecolor{myY2}{RGB}{255,255,205}
\definecolor{myYO}{RGB}{255, 220, 151}
\definecolor{mygreenc}{RGB}{150,50,50}
\usepackage{tensor}
\newcommand{\MYcircle}[2]{
\draw[thick, fill=white] (#1,#2) circle (0.1cm); }
\newcommand{\MYcircleB}[2]{
\draw[thick, fill=black] (#1,#2) circle (0.1cm); }

\usepackage{graphicx} 
\usepackage[normalem]{ulem}

\newcommand{\be}{\begin{equation}}
\newcommand{\ee}{\end{equation}}

\DeclareMathOperator{\aone}{\alpha_1}
\DeclareMathOperator{\atwo}{\alpha_2}
\DeclareMathOperator{\bone}{\beta_1}
\DeclareMathOperator{\btwo}{\beta_2}
\DeclareMathOperator{\cone}{\gamma_1}
\DeclareMathOperator{\ctwo}{\gamma_2}
\DeclareMathOperator{\done}{\delta_1}
\DeclareMathOperator{\dtwo}{\delta_2}
\newcommand{\1}{\mathbbm{1}}
\newcommand{\mcirc}{\mathbin{\scalerel*{\fullmoon}{G}}}

\newcommand{\pk}[1]{{\color{blue}[Pavel: #1]}}
\newcommand{\gs}[1]{{\color{violet}#1}}

\newcommand{\Noise}[3]{
\draw[ thick, fill=myY2, rounded corners=1pt] (#1-0.25,#3+0.25) rectangle (#2+0.25,#3-0.25);
\node at ({(#1+#2)/2}, {#3}) {$\varepsilon$};
}

\newcommand{\Wgategreen}[2]{
\draw[very thick] (#1-0.5, #2 +0.5) -- (#1+0.5,#2-0.5);
\draw[very thick] (#1-0.5,#2-0.5) -- (#1+0.5,#2+0.5);
\draw[ thick, fill=mygreen, rounded corners=2pt] (#1-0.25,#2+0.25) rectangle (#1+0.25,#2-0.25);
\draw[thick] (#1,#2+0.15) -- (#1+0.15,#2+0.15) -- (#1+0.15,#2);
}
\newcommand{\Wgateyellow}[2]{
\draw[very thick] (#1-0.5, #2 +0.5) -- (#1+0.5,#2-0.5);
\draw[very thick] (#1-0.5,#2-0.5) -- (#1+0.5,#2+0.5);
\draw[ thick, fill=yellow, rounded corners=2pt] (#1-0.25,#2+0.25) rectangle (#1+0.25,#2-0.25);
\draw[thick] (#1,#2+0.15) -- (#1+0.15,#2+0.15) -- (#1+0.15,#2);
}

\newcommand{\Bell}[2]{
\draw[very thick] (#1,#2) .. controls (#1+.1,#2-0.20) and (#1+1-.1,#2-0.20) .. (#1+1,#2);
}

\makeatother
\hypersetup{
	pdftitle={Average-computation benchmarking for local expectation values in digital quantum devices},
	pdfsubject={quantum physics},
	pdfauthor={Flavio Baccari, Pavel Kos, Georgios Styliaris},
	pdfkeywords={quantum computation, quantum physics, benchmarking}}
\begin{document}
\title{Average-computation benchmarking for local expectation values in digital quantum devices}

\newcommand{\mpq}{Max Planck Institute of Quantum Optics, 85748 Garching, Germany}
\newcommand{\mcqst}{Munich Center for Quantum Science and Technology (MCQST), 80799 \ Munich, Germany}
\newcommand{\qtech}{Padua Quantum Technologies Research Center, Università degli Studi di Padova, Italy I-35131, Padova, Italy}

\author{Flavio Baccari}
\thanks{Equal contribution. Authors are listed in alphabetical order.}
\affiliation{Dipartimento di Fisica e Astronomia G. Galilei, Università di Padova, I-35131 Padova, Italy}
\affiliation{\qtech}
\author{Pavel Kos}
\thanks{Equal contribution. Authors are listed in alphabetical order.}
\affiliation{\mpq}
\affiliation{\mcqst}
\author{Georgios Styliaris}
\thanks{Equal contribution. Authors are listed in alphabetical order.}
\affiliation{\mpq}
\affiliation{\mcqst}

\begin{abstract}
As quantum devices progress towards a quantum advantage regime, they become harder to benchmark. A particularly relevant challenge is to assess the quality of the whole computation, beyond testing the performance of each single operation.
Here we introduce a scheme for this task that combines the target computation with variants of it, which, when averaged, allow for classically solvable correlation functions. Importantly, the variants exactly preserve the circuit architecture and depth, without simplifying the gates into a classically-simulable set. The method is based on replacing each gate by an ensemble of similar gates, which when averaged together form space-time channels \href{https://doi.org/10.22331/q-2023-05-24-1020}{[P.~Kos and G.~Styliaris, Quantum 7, 1020 (2023)]}.
We introduce explicit constructions for ensembles producing such channels, all applicable to arbitrary brickwork circuits, and provide a general recipe to find new ones through semidefinite programming.
The resulting average computation retains important information about the original circuit and is able to detect noise beyond a Clifford benchmarking regime. Moreover, we provide evidence that estimating average-computation expectation values requires running only a limited number of different circuit realizations.
\end{abstract}

\maketitle

\emph{Introduction.---} Current quantum platforms allow probing many-body systems beyond sizes that can be comfortably simulated classically~\cite{morvan2024phase,king2024computational}. At the same time, the main limitation in the near term is hardware noise and errors, which may corrupt complex computations~\cite{pan2022solving,aharonov2023polynomial,oh2024classical}.
In response to this limitation, near- and mid-term quantum devices rely extensively on methods to recover the ideal outcome from noisy hardware. These range from error mitigation \cite{temme2017error,li2017efficient,smith2021qubit}, applied in classical postprocessing, to error correction performed during the computation \cite{bluvstein2024logical,google2025quantum}. 
Reliable methods to assess the quality of quantum computations play a crucial dual role in this context. First, they serve to detect the presence of noise, providing valuable feedback that guides the improvement of error correction and mitigation strategies. Second, they help gain confidence in the reliability of the output for computations beyond the classically simulable regime.

Given a quantum computation of interest, assessing the quality of its implementation falls into the broad category of benchmarking and certification \cite{eisert2020quantum}. The simplest approach in this context is to test the error rate of single operations. Methods for this range from full gate tomography \cite{chuang1997prescription,nielsen2021gate} to schemes that extract specific fidelity metrics, such as randomized benchmarking \cite{magesan2011scalable} and variations that allow to estimate correlated errors \cite{harper2020efficient,mckay2020correlated,van2023probabilistic}.
However, testing single operations provides little information about the overall quality of the entire computation, especially in regimes of moderate circuit depth \cite{hines2023demonstrating}. 
At the opposite end of the spectrum, verification and certification schemes check that a given device behaves correctly by testing its performance on carefully crafted computations. They normally rely on additional knowledge about the desired output of the computation and leverage it to check how closely the device reproduces it \cite{boixo2018characterizing,hangleiter2017direct,gheorghiu2019verification,cruz2022preparation}.

A highly desirable complement to previous methods is a benchmarking scheme that can test the quality of the overall computation in a regime that is as close as possible to the algorithm of interest. Such a task becomes particularly challenging in today's computations, where most experiments produce data for which the expected output is not known in advance.
Motivated by this challenge, recent proposals introduced randomized motion-reversal techniques that enable estimation of the average fidelity of arbitrary circuits without prior knowledge of their ideal outputs \cite{proctor2022establishing,proctor2022measuring}. While average fidelity is a powerful global metric, it remains highly desirable to complement it with more task-specific benchmarks. In this work, we focus on a particularly natural objective for quantum simulation experiments: assessing the accuracy of local observables at the circuit output.
Current strategies rely on testing performance on simplified circuits, for which properties of the output can be computed efficiently \cite{ferracin2019accrediting,wang2021scalable,kim2023evidence}.
Two typical choices are: i) circuits of smaller size or depth, which allow for brute force simulation, ii) circuits composed of Clifford operations \cite{gottesman1998heisenberg} or matchgates \cite{jozsa2008matchgates}, which admit efficient classical simulation of both expectation values and measurement sampling at the output \cite{bravyi2016improved,bennink2017unbiased,carrasco2024gaining}.
However, altering the circuit to be executed typically affects the noise behavior, as the latter might strongly depend on the circuit architecture and the specificity of the gates. Hence, performing checks on circuits that significantly differ from the target computations has a high chance of providing wrong quality assessments, unless specific assumptions on the noise model are made \cite{merkel2025cliffordbenchmarkssufficientestimating}.

Here, we propose a new benchmarking scheme that avoids any such simplifications and aims to provide a more informative testing regime applicable to any digital quantum computation. The method is based on \emph{randomizing} the desired computation while keeping intact the executed circuit's architecture, depth, and size. The key property is that, although the individual realizations are not classically simulable, calculating few-body correlation functions of the \emph{average} computation is classically efficient. Importantly, the value of the correlation functions carries information about the target circuit and, in general, remains non-trivial. Our schemes do not require ancilla qubits or measurements, and thus each individual realization closely resembles the target circuit.
Next, we outline the general idea and then propose specific examples of such average-computation benchmarking, based on suitably chosen modifications of 2-qubit gates. 
For instance, each gate might be randomly replaced with one of the four variants of it with equal probability.

\emph{Average-computation benchmarking.---}
We consider the following benchmarking task: given a circuit that one is interested in running on the device, how closely does the actual implementation approximate the ideal result of some observables of interest?  
We seek a method that involves executing circuits sharing the same layout, size, and depth as the original circuit. Moreover, each such circuit should, in general, not admit a trivial classical simulation.

In what follows, we introduce the notion of average-computation benchmarking and argue that it satisfies all of the above requirements.
We take a digital model of computation and define the output of an arbitrary quantum circuit as
\begin{equation}
    \mathcal{U} \ (\ketbra{\psi_0}{\psi_0}) = U_k \ldots U_2 U_1 (\ketbra{\psi_0}{\psi_0}) \, ,
\end{equation}
where $\ket{\psi_0}$ is the initial state and the $U_i ( \cdot ) $ are two-qubit gates\footnote{We always assume no single-qubit gate in the layout by simply merging each of them to the nearest two-qubit gate in the circuit.}. Consider also a collection of observables $\mathcal{O}_{\text{test}}$ one is interested in testing at the output.
Assume now that each gate is part of an ensemble $\{ U_{i,\alpha} \}_\alpha$, chosen independently from the others according to a distribution $p_i(U_{i,\alpha})$ and consider the channel resulting from the average over the whole ensemble, namely $\mathcal{E}_i (\cdot ) = \sum_\alpha p_i(U_{i,\alpha}) U_{i,\alpha} ( \cdot ) U_{i,\alpha}^\dagger$.
By fixing a sample over gate realizations, which we call a round, and gathering data from the output of the corresponding circuits, one can estimate the expectation values of interest
\begin{equation}
\langle \mathcal{O}_{\text{test}} \rangle_{\mathcal{U}^{[r]}} = \Tr (  \mathcal{O}_{\text{test}}  \, \, \mathcal{U}^{[r]} \ (\ketbra{\psi_0}{\psi_0}) )   \, 
\end{equation}
where $\mathcal{U}^{[r]}$ is the circuit evolution on the $r^{\text{th}}$ round. This is defined by the two-qubit unitaries $U^{[r]}_{k,\alpha_k} \ldots U^{[r]}_{2,\alpha_2} U^{[r]}_{1,\alpha_1}$, where the $\alpha$ variables represent the fixed outcome of the random sampling for the given round and we also adopted the shorthand notation $\langle \mathcal{O}_{\text{test}} \rangle = \lbrace  \langle O_1 \rangle ,  \langle O_2 \rangle,  \ldots \rbrace$. We then define the results of the average computation to be $\mathbb{E}_{\mathcal{U} \sim p} [\langle \mathcal{O}_{\text{test}} \rangle_{\mathcal{U}}]$.

The main idea behind average-computation benchmarking is that the same average can be computed in two different ways. On the quantum computer side, one can estimate the average via the mean over rounds
\begin{equation}
\langle \mathcal{O}_{\text{test}} \rangle_{\text{avg}}^Q = \frac{1}{R} \sum_{r = 1}^R    \langle \mathcal{O}_{\text{test}} \rangle_{\mathcal{U}^{[r]}} \, .
\end{equation}
On the classical side, we can exploit the linearity of the average to rewrite
\begin{align}
\langle \mathcal{O}_{\text{test}} \rangle_{\text{avg}}^C & =  \Tr (  \mathcal{O}_{\text{test}}  \, \, \mathbb{E}_{\mathcal{U} \sim p} [\mathcal{U}] \ (\ketbra{\psi_0}{\psi_0}) ) \nonumber \\
& =  \Tr (  \mathcal{O}_{\text{test}} \,   \mathcal{E}_{\mathrm{avg}} \ (\ketbra{\psi_0}{\psi_0}) )  \, ,
\end{align}
where $\mathcal{E}_{\mathrm{avg}} \ (\ketbra{\psi_0}{\psi_0}) = \mathcal{E}_k \ldots \mathcal{E}_2 \mathcal{E}_1 (\ketbra{\psi_0}{\psi_0})$  represents the average computation as a non-unitary evolution defined by the concatenation of the average channels $\mathcal{E}_i$.
Note that, to form the individual channels, we used the fact that $p_i$ and $p_j$ are independent if $i \ne j$.
We claim that, for certain testing sets $\mathcal{O}_{\text{test}}$, the averaging ensemble $p(U_{i,\alpha})$ can be chosen so that evaluating the average-computation expectation values $\langle \mathcal{O}_{\text{test}} \rangle_{\text{avg}}^C$ can be done with an efficient classical algorithm. Remarkably, this remains true even for cases in which 
none of the expectation values $\langle \mathcal{O}_{\text{test}} \rangle_{\mathcal{U}^{[r]}}$ for the single circuit runs can be computed efficiently beyond evaluating them on the quantum device.
In what follows, we illustrate concrete examples of ensembles that lead to an efficient classical computation of the average expectation values. For simplicity, we particularize to 1D brickwall circuit layouts and testing set $\mathcal{O}_{\text{test}}$ involving only few-body correlators. However, our method can be straightforwardly adapted to higher spatial dimensions, as illustrated in the End Matter. 
Even these simple correlators are already believed to be difficult to compute for an arbitrary polynomial-sized quantum circuit, with only a few special exceptions \cite{gottesman1998heisenberg,valiant2002quantum,terhal2002classical,bertini2019exact}. 

\emph{Space-time averaging.---} 
As an additional requirement, we want to define benchmarking examples whose average correlations retain some information about the single instances of the ensemble. This is to be contrasted, for instance, to taking Haar-random two-qubit unitaries $U_i$.
In that case, the average value of any correlation function vanishes, independently of the original circuit to be benchmarked.
We pursue a more informative alternative by defining ensembles where the average channels, $\mathcal{E}_i$, are space-time channels~\cite{kos2023circuits}. These remain valid quantum channels (i.e., completely-positive and trace-preserving) upon switching the roles of space and time and lead to classically efficiently computable correlators.

Space-time channels subsequently play an important role in our scheme, so we recall their definition and some basic properties.
For convenience, we will use vectorization, where we map operators to vectors using $\ket{m} \! \bra{n} \xmapsto{{\rm vec}}  \ket{m} \otimes \ket{n} \,$~\cite{kos2023circuits}. For instance, the Hilbert–Schmidt normalized identity operator is mapped to a Bell pair
\begin{align}
 \frac{1}{\sqrt{2}}  \ket{\1_2} = \frac{1}{\sqrt{2}}  \sum_{r=0}^1 \ket{r} \otimes\ket{r}=\ket{\mcirc}=
\begin{tikzpicture}[baseline={([yshift=-0.2ex]current bounding box.center)}, scale=.7]
\draw[thick] (0,0) -- (0,0.6);
\draw[thick, fill=white] (0,0) circle (0.1cm); 
\end{tikzpicture}\, .
\label{eq:Id}
\end{align}
After vectorizing density matrices, a 2-qubit channel (superoperator) $\mathcal E$ with Kraus operators $K_\alpha$ is mapped to an operator
\begin{align}
    \mathcal E \xmapsto{{\rm vec}}  \sum_\alpha K_\alpha \otimes K_\alpha^* =
    \begin{tikzpicture}[baseline={([yshift=-0.5ex]current bounding box.center)}, scale=.7]
\Wgategreen{0}{0}
\end{tikzpicture}
\, ,
\end{align}
where we also introduced  usual graphical tensor network  notation~\cite{cirac2021matrix}.
Using this notation, trace preservation and unitality are denoted as
\begin{align}\label{eq:timeuni}
 \begin{tikzpicture}[baseline=(current  bounding  box.center), scale=.7]
\Wgategreen{0}{0}
\draw[thick, fill=white] (0-0.5,0.5) circle (0.1cm); 
\draw[thick, fill=white] (0.5,0.5) circle (0.1cm); 
\end{tikzpicture}\,  =
\begin{tikzpicture}[baseline=(current  bounding  box.center), scale=.7]
\draw[thick](-.5,-.5)--(-.5,.5);\draw[thick](.5,-.5)--(.5,.5);
\draw[thick, fill=white] (0-0.5,0.5) circle (0.1cm); 
\draw[thick, fill=white] (0.5,0.5) circle (0.1cm); \end{tikzpicture}\;, \quad
\begin{tikzpicture}[baseline=(current  bounding  box.center), scale=.7]
\Wgategreen{0}{0}
\draw[thick, fill=white] (0-0.5,0-0.5) circle (0.1cm); 
\draw[thick, fill=white] (0.5,0-0.5) circle (0.1cm); 
\end{tikzpicture}\,  =
\begin{tikzpicture}[baseline=(current  bounding  box.center), scale=.7]
\draw[thick](-.5,-.5)--(-.5,.5);\draw[thick](.5,-.5)--(.5,.5);
\draw[thick, fill=white] (0-0.5,0-0.5) circle (0.1cm); 
\draw[thick, fill=white] (0.5,0-0.5) circle (0.1cm); \end{tikzpicture}\,,
\end{align}
respectively. Spacetime channels arise by defining the analogous conditions in the spatial direction, that is, left and right space unitality conditions
\begin{align}\label{eq:spaceuni}
 \begin{tikzpicture}[baseline=(current  bounding  box.center), scale=.7]
\Wgategreen{0}{0}
\draw[thick, fill=white] (0-0.5,0-0.5) circle (0.1cm); 
\draw[thick, fill=white] (-0.5,0.5) circle (0.1cm); 
\end{tikzpicture}\,  =
\begin{tikzpicture}[baseline=(current  bounding  box.center), scale=.7]
\draw[thick](-.5,-.5)--(.5,-.5);\draw[thick](-.5,.5)--(.5,.5);
\draw[thick, fill=white] (0-0.5,0-0.5) circle (0.1cm); 
\draw[thick, fill=white] (-0.5,0.5) circle (0.1cm); \end{tikzpicture}
\;, \quad
\begin{tikzpicture}[baseline=(current  bounding  box.center), scale=.7]
\Wgategreen{0}{0}
\draw[thick, fill=white] (0.5,0-0.5) circle (0.1cm); 
\draw[thick, fill=white] (0.5,0.5) circle (0.1cm); 
\end{tikzpicture}\,  =
\begin{tikzpicture}[baseline=(current  bounding  box.center), scale=.7]
\draw[thick](-.5,-.5)--(.5,-.5);\draw[thick](-.5,.5)--(.5,.5);
\draw[thick, fill=white] (0.5,0-0.5) circle (0.1cm); 
\draw[thick, fill=white] (0.5,0.5) circle (0.1cm); \end{tikzpicture}\; .
\end{align}
In particular, we define as \emph{$4$-way space-time channels} those satisfying all four conditions, while as \emph{$3$-way space-time channels} those satisfying \eqref{eq:timeuni} and at least one of \eqref{eq:spaceuni}.

In Ref.~\cite{kos2023circuits}, it was shown how these conditions, for properly-chosen input states, lead to efficiently computable local observables and correlation functions. We present the relevant cases in the End Matter. Here we consider a non-exhaustive list sharing the feature that the average-computation expectation value is generally non-zero and retains dependence on the input circuit. Generally, the structure of correlations is governed by the so-called transfer matrices,
\begin{align}
&\mathcal{M}_+ =
\begin{tikzpicture}[baseline=(current  bounding  box.center), scale=0.55]
\Wgategreen{0}{0}
\MYcircle{.5}{-.5}
\MYcircle{-.5}{.5}
\end{tikzpicture}\; , \quad
&\mathcal{M}_- =
\begin{tikzpicture}[baseline=(current  bounding  box.center), scale=0.55]
\Wgategreen{0}{0}
\MYcircle{.5}{.5}
\MYcircle{-.5}{-.5}
\end{tikzpicture} \;,
\label{eq:MapsM}
\end{align}
which are single-qubit quantum channels that arise from the corresponding average channel $\mathcal E$.
The simplest result concerns both 3-way and 4-way channels a test set $\mathcal{O}_{\text{test}}$ of single-site traceless operators $O^{(T)}$ acting on site $T$, where $T$ is also the circuit depth.
For properly-chosen input states, the average computation reduces to an expression of the form $\langle O^{(T)} \rangle_{\text{avg}} = \bra{O}\prod_{k=1}^T(\mathcal{M}_+)_k\ket{\sigma_X}$, where $k$ runs over average channels in a straight-line path connecting the first input qubit to the output qubit at location $T$. 
Thus, classically evaluating the expression for $\langle O^{(T)} \rangle_{\text{avg}}$ amounts to matrix multiplication; crucially, the dimension of each matrix $\mathcal M$ is only $4$ (for the case of qubits considered here), independent of the system size. 

Average circuits composed of $3$-way channels also allow for efficient classical computation of nearest-neighbour three site observables, $\mathcal{O}_{\text{test}} = \lbrace  O^{(i,i+1,i+2)} \rbrace_{i=0}^{L-3}$.
Lastly, average circuits leading to $4$-way channels allow for efficient classical computation of all 2-body correlators at distance $2T+1$, namely $\mathcal{O}_{\text{test}} = \lbrace O^{(i,i+2T+1)} \rbrace_{i=0}^{L-2T-2}$. We now show explicit choices of ensembles leading, for any unitaries in the original circuit, either to $4$-way or $3$-way average channels.

\emph{4-way averaging.---}
Consider a general parametrisation of the two-qubit unitaries in the circuit
\begin{equation}\label{eq:Ui}
U_i = (W_A \otimes W_B) \exp \left(  i \sum_{\alpha={X,Y,Z}} \theta^{(i)}_\alpha \sigma_\alpha \otimes \sigma_\alpha  \right) (V_A \otimes V_B)    \, ,
\end{equation}
where $W_{A/B}, V_{A/B}$ are arbitrary single-qubit unitary operations~\cite{kraus2001optimal}and $\sigma_\alpha$ are Pauli matrices; note that this simple form is specific to qubits. We rewrite the $\vec{\theta}^{(i)}$ parameters for the two-qubit part as $\vec{\theta}^{(i,++)} = (\frac{\pi}{4} + \delta_x,\frac{\pi}{4} + \delta_y,\theta_z)$. Define the unitary $U_i^{(\pm \pm )}$ as the one that keeps the same single-qubit part of \eqref{eq:Ui} and replaces the vector $\vec{\theta}^{(i)}$ with $\vec{\theta}^{(i,\pm \pm)} = (\frac{\pi}{4} \pm \delta_x,\frac{\pi}{4} \pm \delta_y,\theta_z)$.
A simple ensemble resulting in a $4$-way average channel for each site $i$ is the uniform distribution over the four unitaries $\lbrace U_i^{(++)}, U_i^{(+-)}, U_i^{(-+)}, U_i^{(--)} \rbrace$. This distribution is, by construction, symmetric around the dual-unitary point $\delta_x = \delta_y = 0$~\cite{bertini2019exact}; this property is enough to guarantee that the resulting channel (after averaging) is 4-way unital~\cite{kos2023circuits}.
Notice that the introduced ensemble always includes the original gate, corresponding to the $(++)$ choice.

An alternative ensemble leading to a $4$-way average channel comes from a strategy that resembles Pauli twirling \cite{wallman2016noise}. Instead of replacing gates in the circuit, the ensemble involves applying single-qubit Pauli gates before and after each $U_i$.  In particular, one can directly verify that the average channel defined as
\begin{align}\label{eq:twirling_4way}
\begin{tikzpicture}[baseline={([yshift=-0.9ex]current bounding box.center)},scale=.7]
    \Wgategreen{0}{0}
\end{tikzpicture}
= \frac{1}{4}
\begin{tikzpicture}[
    baseline={([yshift=-0.6ex]current bounding box.center)}, 
    scale=.7,
    circ/.style={
        circle, 
        draw, 
        thick, 
        fill=white, 
        minimum size=1.1em, 
        inner sep=0pt,
        font=\small         
        }
]
    \Wgateyellow{0}{0}
\end{tikzpicture}
+ \frac{1}{12} \sum_{\alpha,\beta={X,Y,Z}} 
\begin{tikzpicture}[
    baseline={([yshift=-0.6ex]current bounding box.center)}, 
    scale=.7,
    circ/.style={
        circle, 
        draw, 
        thick, 
        fill=white, 
        minimum size=1.1em, 
        inner sep=0pt,
        font=\small         
        }
]
    \draw[black, very thick] (.8,-.8) -- (-.8,.8);
    \Wgateyellow{0}{0}
    \path (.8,-.8) -- (-.8,.8)
        node[pos=.2, circ] {$\beta$}
        node[pos=.8, circ] {$\alpha$};
\end{tikzpicture}
\, ,
\end{align}
is $4$-way for any choice of unitary $U_i$, where we defined $\begin{tikzpicture}[
    baseline={([yshift=-0.6ex]current bounding box.center)}, 
    scale=.7,
    circ/.style={
        circle, 
        draw, 
        thick, 
        fill=white, 
        minimum size=1.1em, 
        inner sep=0pt,
        font=\small         
        }
]
    \Wgateyellow{0}{0}
\end{tikzpicture}
= U_i \otimes U_i ^*$ and $\begin{tikzpicture}[
    baseline={([yshift=-0.6ex]current bounding box.center)}, 
    scale=.7,
    circ/.style={
        circle, 
        draw, 
        thick, 
        fill=white, 
        minimum size=1.1em, 
        inner sep=0pt,
        font=\small         
        }
]
    \path (0,0) -- (0,0)
        node[pos=.2, circ] {$\alpha$};
\end{tikzpicture}
= \sigma_\alpha \otimes \sigma_\alpha^*$. Alternatively, the Pauli matrices can be applied to the opposite pair of legs, or one can take an arbitrary convex combination of the two strategies. 

\emph{3-way averaging.---}
A unitary ensemble leading to a $3$-way average channel can be obtained with a simple Pauli twirling procedure applied to one of the output legs of the original unitary $U_i$. More precisely, the average channel resulting from
\begin{align}\label{eq:twirling_3way}
\begin{tikzpicture}[baseline={([yshift=-0.9ex]current bounding box.center)},scale=.7]
    \Wgategreen{0}{0}
\end{tikzpicture}
= \frac{1}{4}
   \sum_{\alpha = \1,X,Y,Z} \begin{tikzpicture}[baseline={([yshift=-0.6ex]current bounding box.center)}, scale=.7,
   circ/.style={
        circle, 
        draw, 
        thick, 
        fill=white, 
        minimum size=1.1em, 
        inner sep=0pt,
        font=\small         
        }]
\draw[black, very thick] (0,0) -- (-.8,.8);
\Wgateyellow{0}{0}
\node[above] at (-0.46,0.02) {\tikz[baseline=(char.base)] \node[draw, circle, thick, fill=white, inner sep=1pt] (char) {$\alpha$};};
\end{tikzpicture}
\, ,
\end{align}
is a $3$-way channel (right-unital) for any input unitary. The averaging procedure corresponds to effectively applying a fully depolarising channel to one of the outputs of each gate in the circuit. 

Another approach is to first simplify the dynamics by adding dephasing channels after each gate, implemented by applying $\sigma_Z$ with 50\% probability.
This simplifies the unitary channel to a reduced channel with only $9$ parameters~\cite{kos2023circuits}. To make the resulting channel 3-way (4-way) unital, one (two) of these parameters needs to vanish in the convex combination of such channels. 
Concretely, we can select a new unitary $U_i'$ such that the target parameter changes sign. The combined channel, formed from an equal mixture of $U_i$ and $U_i'$ together with dephasing, yields a 3-way unital channel. A suitable $U_i'$ can be found systematically using the equations for the aforementioned parameters, which are reported in App.~B1 of Ref.~\cite{kos2021correlations}.

\emph{Relation to the original circuit.---}
Let us elaborate on how the proposed average computations retain information about the original circuit. To make a quantitative assessment, we analyse how closely the Pauli coefficients $(\mathcal{E}_i)_{\beta_1\beta_2,\alpha_1\alpha_2} =  \Tr [\sigma_{\beta_1} \otimes \sigma_{\beta_2} \mathcal{E}_i (\sigma_{\alpha_1} \otimes \sigma_{\alpha_2})]$ of the average channel are related to the coefficients of each original unitary $U_i$.

The first relevant components to study are the set of coefficients forming the transfer matrices of Eq.~\eqref{eq:MapsM}, since they determine non-zero correlations in space-time channels (see End Matter).
Interestingly, in the averaging procedures proposed above, the coefficients of the average channel's transfer matrix are identical to $U_i$'s, or differ only by a rescaling factor $ 1/3 \leq x \leq 1$, which depends on the specific ensemble (see 
Sec.~I of Appendix for details).  
This perspective suggests a more systematic search for averaging strategies by identifying averaging ensembles that preserve the Pauli coefficients of the input unitary as much as possible.

In Sec.~I of Appendix we show how to formulate the problem of searching for a given rescaling supermap as a semidefinite program (SDP). The program allows for searching for averaging strategies that output $4$-way or $3$-way channels with optimal values for the rescaling factors of interest. It can be used to show that no linear supermap can act as a projection on space-time channels such that the resulting channel satisfies \eqref{eq:spaceuni} while keeping the remaining Pauli coefficients $(\mathcal{E}_i)_{\beta_1\beta_2,\alpha_1\alpha_2}$ the same. 
Notice also that not all proposed averaging strategies can be written as supermaps. For instance, the reflection ensemble $U_i^{(\pm \pm )}$ cannot, since the resulting channel does not depend linearly on the input unitary.

Once a supermap is found, it is important to find a way to test whether it can be implemented as an averaging strategy. 
Sec.~II of Appendix provides a linear program \cite{boyd2004convex} to check whether a supermap can be realized as averaging over the application of single-qubit Pauli gates before and after the input unitary. In the same section, we show how the combination of the SDP and the linear program allows us to discover the simple averaging strategies of Eqs.~\eqref{eq:twirling_4way} and \eqref{eq:twirling_3way}.

\begin{figure}
    \centering
    \includegraphics[width=0.45\textwidth]{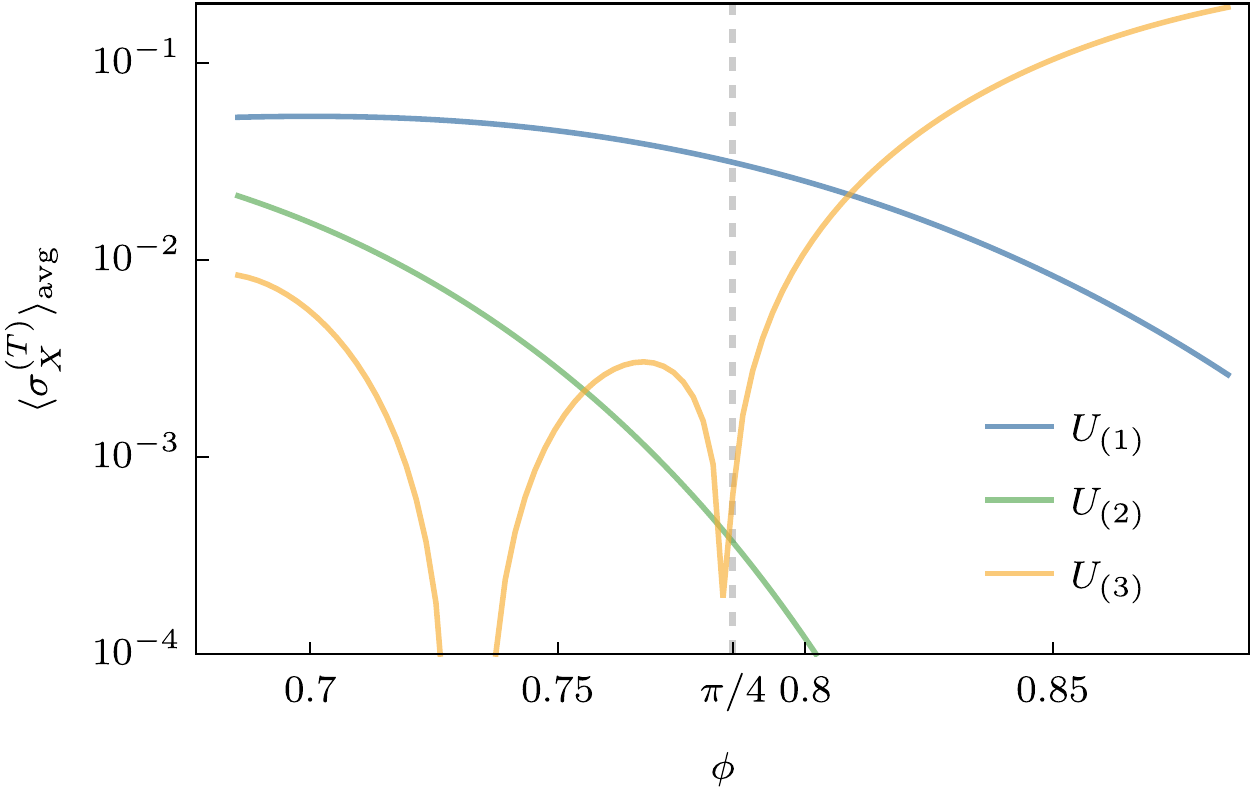}
    \caption{Sensitivity to coherent noise. Expectation values are plotted versus the rotation angle in the imprecise T-gate  $T=\text{diag}(1,e^{i \phi})$ after five layers of the circuit. Dashed line marks the correct value $\phi=\pi/4$.
   The three gates are given in Eq.~\eqref{gateEX}. This demonstrates that our benchmarking scheme can detect this type  of coherent error, in contrast to Clifford benchmarking.
   We take the initial state as explained in the End Matter and operator $\sigma_X$ at position $T$. 
   }
    \label{fig:overRotation}
\end{figure}

\emph{Noise benchmarking.---} Next, we showcase that expectation values on average computations are sensitive to noise in the quantum circuit. Therefore, a discrepancy between the analytical prediction and the experimental value signals the presence of unaccounted-for noise. As an illustrious concrete example of an error that standard Clifford benchmarking would fail to detect, we consider a brickwork circuit, composed of repeating uniformly the gate
\begin{subequations} \label{gateEX}
\begin{align}
    U_{(n)} \!&=\! \text{CNOT}_{12}( u_n\!\otimes \!u_n) \text{CNOT}_{21}( u_n\!\otimes \!u_n )\text{CNOT}_{12} ( u_n\!\otimes \!u_n) 
\end{align}
where we consider three possible options
\begin{align}
    u_1 = H T^2 H T H, 
u_2 = H T^2 H T H T H,
u_3 = H T^2 H T H T,
\end{align}
\end{subequations}
and a concrete noise model, i.e., a consistent but incorrect rotation of the T-gate $T=\text{diag}(1,e^{i \phi})$, where $\phi=\pi/4$ corresponds to no error.
Fig.~\ref{fig:overRotation} shows the analytically computed  expectation values of a local observable versus $\phi$ using our 3-way averaging scheme of Eq.~\eqref{eq:twirling_3way}.

An alternative scenario that fits well with the proposed average-computation benchmarking comes from Pauli noise. Consider a circuit realisation where we assume that every layer is followed by a gate-independent error channel $\mathcal{E}_{\text{noise}} = \sum_{\vec{\alpha}} p_{\vec{\alpha}} \sigma_{\vec{\alpha}} ( \cdot ) \sigma_{\vec{\alpha}} $ which is a mixture of (possibly many-qubit) Pauli errors. Such a model is widely used for error characterisation, error tailoring and error mitigation in current experiments \cite{wallman2016noise,flammia2021pauli,van2023probabilistic,kim2023evidence,rouze2023efficient}.
We show in 
Sec.~IV of Appendix that, under such a noise model, any averaging strategy leading to space-time average channels allows for efficient computation of noisy expectation values as well. In this case, testing the classically-computed values with the ones from the quantum device serves as a benchmark for the given noise model, providing an additional use case for average-computation benchmarking.

\emph{Practical implementation.---} 
We now discuss the sample complexity of estimating the average expectation value $\langle \mathcal{O}_{\text{test}} \rangle_{\text{avg}}^Q$ from the quantum device. Assuming $ \| O_{\text{test}} \| \le 1$, Hoeffding's inequality directly implies that \( M \) independent samples suffice to estimate the expectation up to an additive error $\epsilon \sim 1/\sqrt{M}$ -- independent of the system size $L$.
Building on this, our exact numerical simulations suggest that the variance is, at least in the tested cases, actually much smaller than one, providing evidence that only a few shots are sufficient to estimate the mean expectation value. 
We simulated the first procedure explained after Eq.~\eqref{eq:MapsM} and in the End Matter, i.e., for computing $\langle O^{(x)} \rangle_{\text{avg}}$.
Rather than looking just at the mean value (which can be computed efficiently), we analyzed the entire distribution of the measurement outcomes for a local observable at  sites $x=T$ and $x=T-1$, where in the latter the average expectation value is zero.
The circuit gates were sampled according to a 4-way reflection ensemble $U_i^{(\pm \pm )}$.

\begin{figure}[t]
        \centering
        \includegraphics[width=0.45\textwidth]{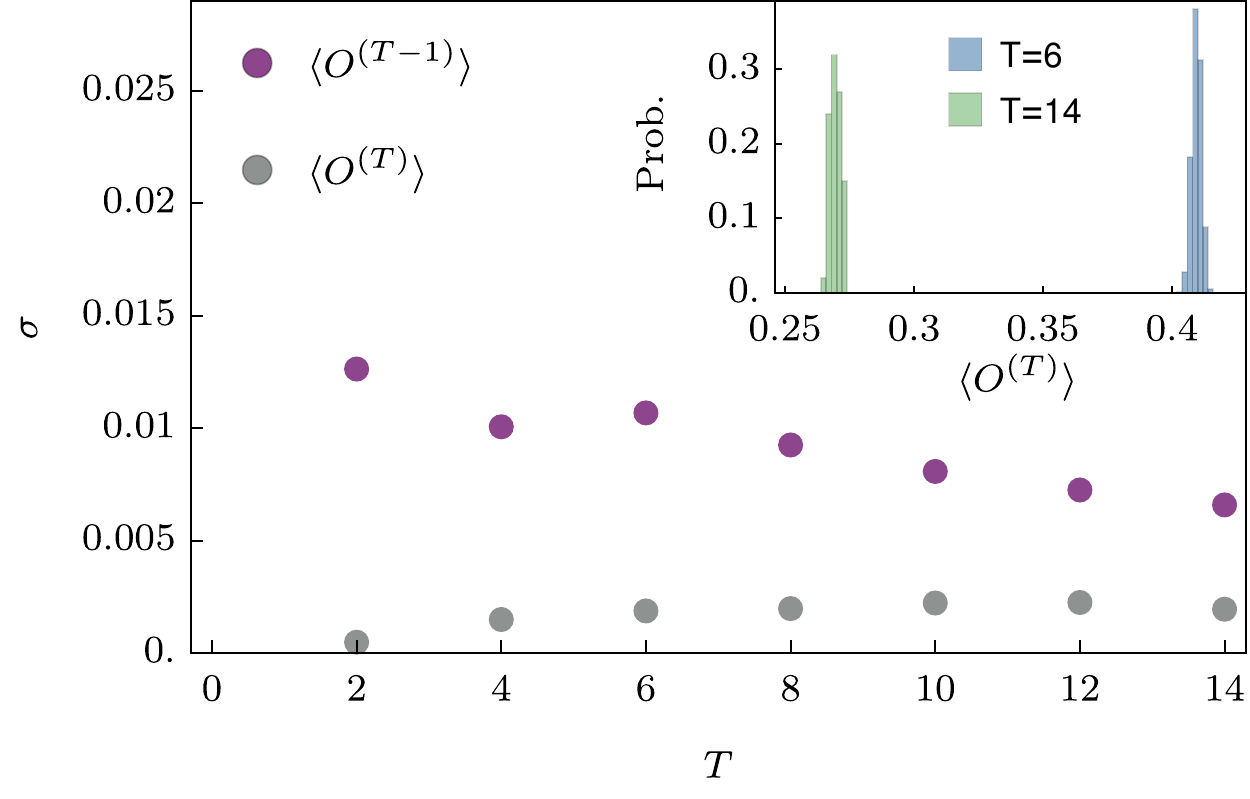}
        \caption{
Main: Standard deviation $\sigma$ of a local observable's expectation value versus circuit depth ($T$), for operators at position $T-1$, where the average is zero, and at position $T$. The standard deviation is calculated over different circuit realizations, where gates are sampled according to a 4-way reflection ensemble $U_i^{(\pm \pm )}$.
Inset: The full distribution of the expectation values $\langle O^{(T)} \rangle$ of different circuit realizations for depths $T=6,14$. The parameters of $U_i^{(\pm \pm )}$ were set to generic values not too far from dual-unitarity, $\theta_z=0.6, \delta_x=\delta_y=0.05$, and we used fixed randomly chosen single-site gates and a fixed operator. The particular choice of single-site gates does not significantly alter the distribution.
 Their specific values are reported in Sec.~III of Appendix.
\label{fig:Histograms}
}
         \label{fig:noiselessHistograms}
    \end{figure}

In Fig. \ref{fig:Histograms} we plot the standard deviation of the histograms versus circuit depth (main panel) and the distribution of the resulting expectation values for each circuit realization 
as a histogram (inset). The results suggest that the standard deviation of the expectation values remains small and does not grow significantly with circuit depth. Consequently, the number of samples required to estimate the average expectation value with constant precision remains small.
These estimated averages can then be compared to the corresponding analytical results for benchmarking purposes.

Secondly, we analyze how the correlations used for benchmarking depend on the original circuit. As discussed before, correlations in our average computations are determined by suitable multiplication of the transfer matrices \eqref{eq:MapsM}, which are inherited from the original unitaries in the circuit. By following the analysis in \cite{kos2023circuits}, the second leading eigenvalues of those matrices determine the speed of decay of the average correlations used for benchmarking. We could not find any example of averaging ensemble leading to a transfer matrix with two degenerate eigenvalues equal to $1$, beyond the trivial case of dual-unitary gates, which would make the initial circuits already solvable. Hence, we typically expect the average correlations $\langle \mathcal{O}_{\text{test}} \rangle_{\text{avg}}^C$ to decay exponentially with the circuit depth, likely requiring an increasing number of samples to benchmark. Averaging strategies that rescale the initial transfer matrix, such as \eqref{eq:twirling_4way}, will increase the decay speed by the rescaling factor. 
This makes average-computation benchmarking mostly suited to low-depth quantum circuits, as those addressed in typical near-term quantum computations.

\emph{Conclusions.---}
In this work, we introduced average-computation benchmarking of quantum devices. 
The method involves replacing each gate in the circuit with one chosen from an ensemble of similar gates. While each individual random circuit likely remains as computationally complex as the original, the expectation values of local observables, when averaged over the ensemble, become classically tractable.

Our scheme offers important advantages over existing methods. It assesses the performance of the entire computation without simplifying the original circuit or altering its size and depth, 
We also demonstrated that the method is sensitive to coherent errors that evade standard Clifford-based protocols and that the number of experimental runs required to estimate the benchmark values is manageable.

Poor agreement with average-computation expectation values provides a direct signature of uncompensated noise in the device. As with other benchmarking schemes, however, passing the test does not strictly speaking guarantee a flawless computation, since certain errors may be suppressed by the averaging procedure. Our method should be thought of as part of a broader set of complementary tests that strengthen confidence in the validity of the executed computation.

Beyond hardware characterization, the benchmarking procedure can directly be used to evaluate the performances of classical simulation algorithms for many-body dynamics and quantum computation.

Generalizing our proposed approach to qudits is straightforward. While the reflection strategy is not easily generalizable due to the lack of a standard decomposition for general qudit gates, our other strategies are readily adaptable. The generalization would also provide an alternative benchmarking scheme for qubits by grouping several sites together to form effective qudits. 

The local correlators used in our examples decay exponentially with circuit depth, making the scheme most suitable for near- and mid-term applications. A promising future direction is to explore benchmarking schemes based on non-decaying signals, such as out-of-time-ordered correlators (OTOCs). This would require an extension of the current framework, as computing OTOCs involves 'two-replica' quantities that are not solvable within the class of space-time channels presented here.

\begin{acknowledgments}
\emph{Acknowledgments.---}
We thank Bruno Bertini and Ignacio Cirac for valuable discussions. The work at MPQ
is partly funded by THEQUCO as part of the Munich Quantum Valley, which is supported by the Bavarian state government with funds from the Hightech Agenda Bayern Plus. P.K. acknowledges financial support from the Alexander von Humboldt Foundation. F.B. acknowledges financial support from the ICSC – Centro Nazionale di Ricerca in High Performance Computing, Big Data and Quantum Computing, funded by European Union – NextGenerationEU and from the European Union’s Horizon Europe research and innovation programme under the Marie Skłodowska-Curie Action for the project No. 101148556 (ENCHANT).  The views and opinions expressed here are solely those of the authors and do not necessarily reflect those of the funding institutions. Neither of the funding institutions can be held responsible for them. 
\end{acknowledgments}

\bibliography{biblio}

\newpage
\onecolumngrid
\appendix
\section*{End Matter}
\twocolumngrid
\section*{Benchmarking schemes} 
Here we detail how 4-way and 3-way unital channels simplify the calculation of certain local expectation values at the output of an average computation, where each unitary in the circuit is replaced by a channel.
We use the diagrammatic folded representation introduced in the \emph{space-time averaging} section (see also Ref.~\cite{kos2023circuits}).

As in the main text, we formulate the quantities to be tested in terms of observables. The schemes can, however, be directly adapted to yield the corresponding reduced state by contracting the same diagrams with the observables removed.

\emph{Initial states.---} Our procedure requires compatible solvable initial states~\cite{kos2023circuits}, or an averaging procedure that transforms the initial states into solvable ones.
States that are compatible with our procedure are solvable matrix product density operators (MPDOs)~\cite{kos2023circuits}. Their defining characteristic is that they are trace-preserving maps from left to right or vice versa, where we group the left (right) physical and auxiliary indices together. 
In this work, we will limit ourselves to a simple example, a product state on sites $0,\dots,L-1$ composed of Bell pairs, $|\Phi^+\rangle = \frac{1}{\sqrt{2}} \left( |00\rangle + |11\rangle \right)$:
\begin{align}
\rho(0)= \rho_{\rm{Bell}}^{\otimes L/2}, \quad \rho_{\rm{Bell}} &= |\Phi^+\rangle \langle \Phi^+|.
\end{align}
The state preparation can be treated as an additional single layer of gates, as compared with the target circuit.

\emph{Two-body correlations: 4-way channels.---} Once a solvable initial state $\rho_0$  is prepared, we can compute physical observables. First we consider the two-body correlations  between local operators $a^{(i)}$ and $b^{(j)}$ at position $i$ and $j$ respectively, captured by two body operator $O^{(i,j)}=a^{(i)}b^{(j)}$. After evolution by $T$ layers   of circuit, they are given by 
$
\langle O^{(i,j)} \rangle_\text{avg}
= 
\text{Tr}\left( O^{(i,j)} \rho(T) \right),
$
where we assume $j>i$. $\rho(T)$ results from a quantum evolution of $T$ layers with a brick-wall circuit composed of local quantum channels.
Graphically, we express the two-point equal time correlation function  as: 
\begin{align}\label{eq:quench2}
&\langle O^{(i,j)} \rangle_\text{avg} \!=\! \begin{tikzpicture}[baseline=(current  bounding  box.center), scale=0.6]
\foreach \i in {0,...,1}{
\Wgategreen{-2}{0+2*\i}\Wgategreen{0}{0+2*\i}\Wgategreen{2}{0+2*\i}\Wgategreen{4}{0+2*\i}\Wgategreen{6}{0+2*\i}
\Wgategreen{-1}{1+2*\i}\Wgategreen{1}{1+2*\i}\Wgategreen{3}{1+2*\i}\Wgategreen{5}{1+2*\i}\Wgategreen{7}{1+2*\i}
}
\foreach \i in {2.5,4.5,6.5,8.5,10.5}
{\Bell{-4+\i}{-.5}}
 \foreach \i in {-1,...,8}{
 \draw[thick, fill=white] (\i-0.5,4-0.5) circle (0.1cm); 
}
\MYcircleB{7.5}{3.5}
\MYcircleB{-1.5}{3.5}
\Text[x=-1.5,y=4.0]{$a^{(i)}$}
\Text[x=7.5,y=4.0]{$b^{(j)} $}
\end{tikzpicture}.
\end{align}

In the presence of left and right unitality and trace preservation, which is the case for 4-way channels, the diagram simplifies considerably.
 In particular, for $j > i+2T+1$ and our specific solvable initial states, the correlations are equal to zero.
Instead, for $j = i+2T+1$ the diagram can be simplified to~\cite{kos2023circuits}:
\be
\langle O^{(i,i+2T+1)} \rangle_\text{avg} \!=\!
\begin{tikzpicture}[baseline=(current  bounding  box.center), scale=0.6]
\Wgategreen{2}{0}\Wgategreen{4}{0}
\Wgategreen{-1}{3}\Wgategreen{7}{3}
\Wgategreen{0}{2}\Wgategreen{6}{2}
\Wgategreen{1}{1}\Wgategreen{5}{1}
%
\Bell{2.5}{-.5}
\foreach \i in {0,...,3}{
\MYcircle{\i-.5}{3.5-\i}
\MYcircle{\i-.5-1}{3.5-\i-1}
\MYcircle{\i+4-.5}{0.5+\i}
\MYcircle{\i+4-.5+1}{0.5+\i-1}
}
\MYcircleB{7.5}{3.5}
\MYcircleB{-1.5}{3.5}\Text[x=-1.5,y=4.0]{$a$}
\Text[x=7.5,y=4.0]{$b$}
\end{tikzpicture}.
\label{eq:quench5}
\ee
First, we simplified from the top using trace preservation, which created backward light cones from the local operators. Then we use left and right unitality to simplify the diagram from the sides. For $j < i+2T+1$ the diagram is almost the same, except that the initial Bell pair is changed to  $(\text{vec}\; \mathcal{E}) \ket{\mcirc\mcirc}$. In the presence of unitality, which is a property of all our average channels, this also simplifies to zero.  

Eq.~\eqref{eq:quench5} is an effective one-dimensional  diagram that can be contracted efficiently.
The final expression can be written using the transfer matrices from Eq.~\eqref{eq:MapsM} as 
\begin{align}
  \langle O^{(i,i+2T+1)} \rangle_\text{avg}\!=\!\langle a b |  \!\prod_{k=1}^T\!(\mathcal{M}_-)_k \!\otimes\! (\mathcal{M}_+)_k \!\ket{\rho_{\rm{Bell}}}\!,\!
\end{align}
where we added the subscript $k$ to denote that $\mathcal{M}_\pm$ may be different in different layers.

\emph{Nearest-neighbour k-body correlations: 3-way channels.---} The simplification from the previous section cannot be achieved without both left and right unitality. Nevertheless, the case of 3-way channels is still interesting and tractable. Let us assume the channel is unital, trace-preserving, and right-unital.
In this scenario, we can consider the expectation value of  an observable acting on three consecutive sites: $\langle O^{(i,i+1,i+2)}\rangle_\text{avg}=\tr ( O^{(i,i+1,i+2)}\rho(T) )$. For this calculation, we simplify the corresponding diagram from the top (using trace preservation) and from the right (using right unitality),  leading to the following diagram:
\be
\langle O^{(i,i+1,i+2)}\rangle_\text{avg} =
\begin{tikzpicture}[baseline=(current  bounding  box.center), scale=0.6]
\Wgategreen{5}{3}\Wgategreen{3}{3}
\Wgategreen{4}{2}\Wgategreen{2}{2}
\Wgategreen{3}{1}\Wgategreen{1}{1}
\Wgategreen{2}{0}\Wgategreen{0}{0}
\foreach \i in {0,...,2}{
\MYcircle{\i-.5+1}{1.5+\i}
\MYcircle{\i-.5}{1.5+\i-1}
\MYcircle{\i+4-.5+1-1}{0.5+\i}
}
\MYcircleB{5.5}{3.5}
\MYcircleB{4.5}{3.5}
\MYcircleB{3.5}{3.5}
\Bell{0.5}{-.5}
\Bell{2.5}{-.5}
\Bell{-1.5}{-.5}
\MYcircle{3.5}{-0.5}
\MYcircle{-1.5}{-0.5}
\end{tikzpicture},
\label{eq:quench7}
\ee
where the black dots denote the position of the observable.
This effectively one-dimensional diagram can be efficiently contracted using repeated actions of two-qubit quantum channels 
\begin{align}
&\mathcal{M}_+^2 =
\begin{tikzpicture}[baseline=(current  bounding  box.center), scale=0.55]
\Wgategreen{0}{0}
\Wgategreen{-1}{1}
\MYcircle{.5}{-.5}
\MYcircle{-1.5}{1.5}
\end{tikzpicture}.
\end{align}
The final expression reads
\begin{align}
    \langle O^{(i,i+1,i+2)}\rangle_\text{avg}\!=\!
    \bra{O} \!(\1 \otimes \mathcal{E}_L) \!\prod_{k=1}^{T-1} (\mathcal{M}_+^2)_k (\mathcal{M}_R \!\otimes\! \1) \!\ket{\rho_{\rm{Bell}}},
\end{align}
where $O$ is a three-site operator, and 
\begin{align}
    \mathcal{M}_R = \bra{\mcirc}_L (\text{vec} \; \mathcal{E})_1 \ket{\mcirc}_L \quad
    \mathcal{E}_L = (\text{vec} \;\mathcal{E})_T \ket{\mcirc}_R.
\end{align} 
 $(\text{vec} \; \mathcal{E})_i$ denotes the vectorized local channel in layer $i$ and $L,R$ denote the left or right site.
Notice that a similar procedure can be used to simplify the calculation of any nearest-neighbour $k$-body correlations $\langle O^{(i,\ldots,i+k-1)}\rangle_\text{avg}=\tr ( O^{(i,\ldots,i+k-1)}\rho(T) )$. The generalised expression will involve transfer matrices acting on more and more qubits, hence resulting in a contraction effort that scales exponentially with $k$. 
Hence, the computation of correlations will remain efficient, as long as one considers a finite $k$. 

\emph{Single-body expectation values: 3-way channels and non-translational invariant initial state.---}
An alternative approach is to evaluate quantities that mimic infinite-temperature correlation functions. This approach, proposed in~\cite{fischer2024dynamicalsimulationsmanybodyquantum}, works for both $4$-way and the aforementioned $3$-way unital channels. It uses an initial state of plus state $\ket{+}=1/\sqrt{2}(\ket{0}+\ket{1})$ and Bell pairs under open boundary conditions:
\be
\rho(0) = \ket{\Psi_0} \bra{\Psi_0},  \quad  \ket{\Psi_0} = | + \Phi^+ \Phi^+ \dots \rangle.
\ee

We then consider the expectation value of a single-site observable at site $T$ (first site is at position $0$)\footnote{For other positions, the expectation values vanish.}, namely $\langle O^{(T)} \rangle_{\text{avg}} = \text{Tr}\left( O^{(T)} \rho(T) \right)$. The corresponding diagram can be simplified from the top (using trace-preservation) and from the right (using right-unitality), yielding:
\be
\langle O^{(T)} \rangle_{\text{avg}} =
\begin{tikzpicture}[baseline=(current  bounding  box.center), scale=0.6]
\Wgategreen{3}{3}
\Wgategreen{2}{2}
\Wgategreen{1}{1}
\Wgategreen{0}{0}
\foreach \i in {0,...,2}{
\MYcircle{\i-.5+1}{1.5+\i}
\MYcircle{\i-.5}{1.5+\i-1}
\MYcircle{\i+2-.5+1-1}{0.5+\i}
}
\MYcircleB{3.5}{3.5}
\Bell{0.5}{-.5}
\MYcircle{1.5}{-0.5}
\Text[x=-0.5,y=-.75]{$+$}
\Text[x=3.5,y=4.0]{$O$}
\end{tikzpicture}.
\label{eq:quench6}
\ee
This expression matches the infinite temperature  space-time two-point correlation function $\text{Tr}\left(  O^{(T)}(\sigma_X)^{(0)}(T) \right)= \bra{O}\prod_{k=1}^T(\mathcal{M}_+)_k\ket{\sigma_X}$~\cite{bertini2019exact,fischer2024dynamicalsimulationsmanybodyquantum}. In contrast to $2T$ single-site transfer matrices in Eq.~\eqref{eq:quench5}, it uses only $T$ of them. As a result, the correlations decay to a similar level at twice the circuit depth. This slower decay is beneficial for benchmarking purposes, as the signal persists longer and is easier to measure in deeper circuits.

\emph{More general initial states.---} More general initial states can also be accommodated. This can be achieved in two ways: a) by taking convex combinations of different initial states to form solvable states, or b) by applying a layer of local channels that transforms the initial state into a solvable MPDO. A trivial example of the latter is a layer of totally depolarizing channels.

As an illustrative example of method a), we consider an initial pure state described by an injective MPS with two-dimensional physical and bond dimensions. Here, we can think of the MPS tensor as a unitary gate on two qubits. Then we can use the reflection procedure developed for unitary gates (4-way averaging in the main text) to construct a solvable MPDO. When the dimensions of the auxiliary and physical spaces are different, finding these ``reflections'' is more difficult, but we expect that a modification of the SDP procedure outlined in Sec.~I of Appendix for this task is possible.

\emph{Benchmarking schemes in 2+1D.---} 
These ideas can be extended to two-dimensional systems. This requires generalizing dual unitarity to 2D, known as ternary unitarity~\cite{milbradt2023ternary,suzuki2022computational,jonay2021triunitarity},  and combining it with the concept of space channels.

In two dimensions, we first focus on brick-wall evolution, which is composed of plaquette gates acting on four sites arranged in two layers~\cite{milbradt2023ternary}. These gates are ternary unitary if, in addition to normal unitarity, they remain unitary when either of the two spatial dimensions is reinterpreted as the time direction.

Analogous to the 1D case, we can generalize these ideas to local quantum channels and construct plaquette channels that satisfy up to six  unitality conditions. These conditions are sufficient to simplify the correlation functions, ensuring they are nonzero only along the light rays, exactly as for the unitary case in Ref.~\cite{milbradt2023ternary}.

A randomising procedure leading to average $6$-way unital channels can be easily applied to a concrete circuit architecture on a square lattice. The circuit is composed of successive brickwall layers of the form 
\begin{align}
\mathcal{U} = U_{oo,\text{vert}}U_{oo,\text{horz}}U_{ee,\text{vert}}U_{ee,\text{horz}},
\end{align}
where $U_{ee,\text{horz}}$ ($U_{oo,\text{horz}}$) are tensor products of arbitrary two-qubit unitaries acting on sites with even (odd) valued $x$ coordinate
and their right neighbors, and $U_{ee,\text{vert}}$ ($U_{oo,\text{vert}}$) between even (odd) valued $y$ coordinate with their upper
neighbors. Then, applying the reflection strategy to each of the gates results in 4-way unital two-qubit channels. Combining vertical and horizontal interactions of even (odd) layer into plaquette channels, such as local channels arising from $U_{oo}=U_{oo,\text{vert}}U_{oo,\text{horz}}$, 
results in 6-way unital channels.


We believe that more elaborate schemes are possible, but they are beyond the scope of this work.

\clearpage
\appendix

\setcounter{equation}{0}
\setcounter{figure}{0}
\setcounter{table}{0}
\makeatletter
\renewcommand{\theequation}{S\arabic{equation}}
\renewcommand{\thefigure}{S\arabic{figure}}

\onecolumngrid
\section*{Appendix}
\appendix

\section{Systematic search of $4$-way and $3$-way averaging strategies}\label{app:supermap}

We can look for strategies mapping any unitary to a $3$- or $4$-way channel with the general framework of supermaps \cite{chiribella2008transforming}. Let us quickly revise the main technical ingredients required to define supermaps. A supermap is any linear operation transforming a quantum channel mapping states from $\mathcal{H}_{\text{in}}$ to $\mathcal{H}_{\text{out}}$ into another quantum channel mapping states from $\mathcal{K}_{\text{in}}$ to $\mathcal{K}_{\text{out}}$. From now on, we will always assume that the dimensions of the four Hilbert spaces coincide. Supermaps can be seen as a transformation $\tilde{S}$ from the Choi state $\Lambda$ of the input channel to the Choi state of the output channel $\Lambda' = \tilde{S}(\Lambda)$.
In a similar way to Choi states for channels, we can associate with any supermap a positive-semidefinite operator 
\begin{equation}\label{eq_app:superchoi}
    S = \tilde{S}_{\mathcal{K}} \otimes \mathbbm{1}_{\mathcal{H}} ( \ketbra{\Phi^+}{\Phi^+}_{\mathcal{H}\mathcal{K}} ) \, ,
\end{equation}
where $\ket{\Phi^+} = \sum_{i} \ket{i}_{\mathcal{H}_{\text{in}} \otimes \mathcal{H}_{\text{out}}} \ket{i}_{\mathcal{K}_{\text{in}} \otimes \mathcal{K}_{\text{out}}}$. The output Choi state is obtained as
\begin{equation}\label{eq_app:outputChoi}
    \tilde{S} (\Lambda) = \text{Tr}_{\mathcal{H}}[(\mathbbm{1}\otimes \Lambda^T) S] \, .
\end{equation}
In our setting, we will consider supermaps acting on the Choi states corresponding to two-qubit unitaries. For future convenience, we express the supermap action by expanding both the Choi state and the $S$ operator in the Pauli basis. We have
\begin{equation}
\Lambda = \sum_{\aone, \atwo , \bone , \btwo }     \Lambda_{\aone \atwo , \btwo \btwo} \, \,  \sigma_{\aone}^T  \sigma_{\atwo}^T \sigma_{\bone}  \sigma_{\btwo} ,      
\end{equation}
where notice that $\Lambda_{\aone, \atwo , \bone , \btwo }$ correspond to the Pauli coefficient of the input unitary, namely $\Lambda_{\aone, \atwo , \bone , \btwo }= \Tr(\sigma_{\bone}  \sigma_{\btwo} U (\sigma_{\alpha_1}  \sigma_{\alpha_2}) U^\dagger)$. Then \eqref{eq_app:outputChoi} can be rewritten as
\begin{equation}
    \Lambda'_{\aone \alpha_2 , \bone \btwo} = \sum_{\cone, \ctwo , \done , \dtwo} S_{\aone \atwo , \bone  \btwo }^{\cone \ctwo , \done  \dtwo} \Lambda_{\cone \ctwo , \done  \dtwo} ,
\end{equation}
where
\begin{equation}
    S = \sum_{\aone, \atwo , \bone , \btwo ,\cone, \ctwo , \done , \dtwo} S_{\aone \atwo , \bone  \btwo }^{\cone \ctwo , \done  \dtwo} \sigma_{\aone}^T  \sigma_{\atwo}^T \sigma_{\bone}^T  \sigma_{\btwo}^T \sigma_{\cone}  \sigma_{\ctwo} \sigma_{\done}  \sigma_{\dtwo} \, .
\end{equation}
As shown in \cite{chiribella2008transforming}, any supermap defined by a positive operator admits a physical realisation in terms of unitaries applied before and after the two-qubit unitary, acting on the two qubits involved and potentially additional ancillary qubits. For our purposes, we need the output channel $\Lambda'$ to be a $3$- or $4$-way channel. Hence, for a $3$-way channel it needs to satisfy the conditions $\Lambda'_{\aone \mathbbm{1} , \bone \mathbbm{1}} = 0$ for all $\aone, \bone \neq \mathbbm{1}$. While for a $4$-way channel it needs also to satisfy $\Lambda'_{ \mathbbm{1} \atwo , \mathbbm{1} \btwo} = 0$ for all $\atwo, \btwo \neq \mathbbm{1}$. We will thus restrict our search to supermaps satisfying
\begin{equation}\label{eq_app:3way}
    S_{\aone \mathbbm{1} , \bone \mathbbm{1}}^{\cone \ctwo , \done  \dtwo} = 0 \quad \forall \, \, \cone \ctwo , \done  \dtwo  \, \text{and} \, \aone, \bone \neq \mathbbm{1} \, ,
\end{equation} 
and, for $4$-way channels, also
\begin{equation}\label{eq_app:4way}
    S_{\mathbbm{1} \atwo , \mathbbm{1} \btwo}^{\cone \ctwo , \done  \dtwo} = 0 \quad \forall \, \, \cone \ctwo , \done  \dtwo  \, \text{and} \, \atwo, \btwo \neq \mathbbm{1} \, .
\end{equation} 
As an additional constraint, we look for supermaps that preserve as much information as possible about the original input unitary. Mathematically, this translates into the condition
\begin{equation}\label{eq_app:rescaling}
S_{\aone \atwo , \bone  \btwo }^{\cone \ctwo , \done  \dtwo}  = \delta_{\aone \cone}  \delta_{\atwo \ctwo}  \delta_{\bone \done} \delta_{\btwo \dtwo}   x_{\aone \atwo , \bone  \btwo } \, .
\end{equation}
We can interpret any supermap of the form \eqref{eq_app:rescaling} as almost an identity supermap, where any Pauli coefficient of the output channel is proportional to the corresponding coefficient of the input unitary, up to a rescaling factor $x_{\aone \atwo , \bone  \btwo }$. For this kind of supermaps, the conditions \eqref{eq_app:3way} and \eqref{eq_app:4way} correspond to setting the rescaling factors $x_{\aone \mathbbm{1} , \bone \mathbbm{1}}$ and $x_{\mathbbm{1} \atwo , \mathbbm{1} \btwo}$ to zero.  
We stress that supermaps of the above form possess other desirable features by construction. First, the output channel is trace preserving whenever the input channel is trace preserving as well (which is the case for the unitaries in our circuits). Similarly, the output channel is unital whenever the input channel is unital (which is also the case for the unitaries in a circuit). This is a desirable property for us, since we ideally look for supermaps corresponding to a mixture of unitary operations, which are unital by construction.

Lastly, out of all the terms of the input unitary, we look for supermaps that preserve the right and left transfer matrix as much as possible. That is, we look for the rescaling factors $x_{\mathbbm{1} \atwo \bone \mathbbm{1}}$ and $x_{\aone \mathbbm{1}  \mathbbm{1} \btwo}$ to be all the same for different $\atwo, \bone$ and  $\aone, \btwo$. This can be turned into a semidefinite optimisation problem \cite{boyd2004convex}. For instance, the case for a supermap generating $4$-way channels becomes:
\begin{align}
\label{eq_app:4waySDP}
    x_{\text{TM}}^{\text{4way}} = \max_{\vec{x}} &  \, \,  ( x_{\text{RTM}} + x_{\text{LTM}} ) \\ \nonumber
    \text{s.t. } & S = \sum_{\aone, \atwo , \bone , \btwo} x_{\aone \atwo , \bone  \btwo } \sigma_{\aone}^T  \sigma_{\atwo}^T \sigma_{\bone}^T  \sigma_{\btwo}^T \sigma_{\aone}  \sigma_{\atwo} \sigma_{\bone}  \sigma_{\btwo} \succeq 0 \\ \nonumber
    & x_{\aone \mathbbm{1} , \bone \mathbbm{1}} = 0 \quad \forall \aone, \bone \neq \mathbbm{1}  \nonumber \\
    & x_{\mathbbm{1} \atwo , \mathbbm{1} \btwo} = 0 \quad \forall \atwo, \btwo \neq \mathbbm{1}  \nonumber \\
    & x_{\mathbbm{1} \atwo \bone \mathbbm{1}} = x_{\text{LTM}} \quad \forall \atwo, \bone \neq \mathbbm{1}  \nonumber \\
    & x_{\aone \mathbbm{1}  \mathbbm{1} \btwo} = x_{\text{RTM}} \quad \forall \aone, \btwo \neq \mathbbm{1}  \nonumber \\
    & x_{\mathbbm{1} \mathbbm{1} \mathbbm{1} \mathbbm{1}} = 1 \, , \nonumber
\end{align}
where the last condition is imposed to normalise the output Choi state.
Similarly, looking for a supermap generating $3$-way channels can be cast as
\begin{align}
\label{eq_app:3waySDP}
    x_{\text{TM}}^{\text{3way}} = \max_{\vec{x}} &  \, \,  x_{\text{RTM}} \\ \nonumber
    \text{s.t. } & S = \sum_{\aone, \atwo , \bone , \btwo} x_{\aone \atwo , \bone  \btwo } \sigma_{\aone}^T  \sigma_{\atwo}^T \sigma_{\bone}^T  \sigma_{\btwo}^T \sigma_{\aone}  \sigma_{\atwo} \sigma_{\bone}  \sigma_{\btwo} \succeq 0 \\ \nonumber
    & x_{\mathbbm{1} \atwo , \mathbbm{1} \btwo} = 0 \quad \forall \atwo, \btwo \neq \mathbbm{1}  \nonumber \\
    & x_{\aone \mathbbm{1}  \mathbbm{1} \btwo} = x_{\text{RTM}} \quad \forall \aone, \btwo \neq \mathbbm{1}  \nonumber \\
    & x_{\mathbbm{1} \mathbbm{1} \mathbbm{1} \mathbbm{1}} = 1 \, . \nonumber
\end{align}
Notice that we do not impose one of the space unitality conditions, and that we aim at maximizing only the right transfer matrix, since it is the only one contributing to the computation of the solvable correlations. Interestingly, the optimisation problem \eqref{eq_app:3waySDP} admits a solution preserving entirely the transfer matrix (i.e. $x_{\text{TM}}^{\text{3way}} = 1$), while \eqref{eq_app:4waySDP} does not. The best total rescaling factor coming from a $4$-way supermap amounts to $x_{\text{TM}}^{\text{4way}} = 4/3$: 
that is, one of the transfer matrices can be completely preserved, while the other has to be rescaled by $1/3$. Let us analyse the above solutions in detail.

\textit{$3$-way case}. We start from the optimal solutions of \eqref{eq_app:3waySDP}, which corresponds to the supermap defined by the following rescaling parameters
\begin{equation}
\begin{array}{c|cccc}\label{eq_app:3waysol}
x_{\cdot,\cdot} & \beta_1 \beta_2 & \beta_1 \mathbbm{1} & \mathbbm{1} \beta_2 & \mathbbm{1}\mathbbm{1} \\
\hline
\alpha_1 \alpha_2 & 0 & 1 & 0 & 1 \\
\alpha_1 \mathbbm{1} & 0 & 1 & 1 & 1 \\
\mathbbm{1} \alpha_2 & 0 & 0 & 0 & 1 \\
\mathbbm{1}\mathbbm{1} & 0 & 1 & 0  & 1 \\
\end{array}
\end{equation}
where we have joined together all the indices $\alpha_i,\beta_j = X,Y,Z$. This means, for instance, that $x_{\aone \atwo , \bone  \btwo} = 0$ for all $\alpha_i,\beta_j = X,Y,Z$. One can check that the above supermap corresponds to the twirling strategy described in Eq.~\eqref{eq:twirling_3way}.

\textit{$4$-way case}. We found a continuous family of optimal solution of \eqref{eq_app:4waySDP}, parametrized by a coefficient $0 \leq \lambda \leq 1$. The corresponding rescaling parameters read as follows 
\begin{equation}
\begin{array}{c|cccc}\label{eq_app:4waysol}
x_{\cdot,\cdot} & \beta_1 \beta_2 & \beta_1 \mathbbm{1} & \mathbbm{1} \beta_2 & \mathbbm{1}\mathbbm{1} \\
\hline
\alpha_1 \alpha_2 & \frac{1}{3} & \frac{1 - \lambda}{3} & \frac{\lambda}{3} & 0 \\
\alpha_1 \mathbbm{1} & \frac{\lambda}{3} & 0 & \frac{1 + 2\lambda}{3} & \lambda \\
\mathbbm{1} \alpha_2 & \frac{1 - \lambda}{3} & 1 - \frac{2\lambda}{3} & 0 & 1 - \lambda \\
\mathbbm{1}\mathbbm{1} & 0 & 1 - \lambda &  \lambda  & 1 \\
\end{array}
\end{equation}
where we have again joined together all the indices $\alpha_i,\beta_j = X,Y,Z$.
Solutions of this type are convex combinations of two extremal solutions, corresponding to the parameters $\lambda = 0,1$. The extremal solutions are supermaps, which preserve entirely the right or the left transfer matrix, respectively, while rescaling the other one by $1/3$.

\section{Representing a supermap as a convex combination of input-output Pauli gates}\label{sec_app:Paulisupermap}

The most general supermap can be realised by coupling the system with an ancillary space and applying two system-ancilla unitary operations before and after the application of the channel acting on the system. However, we know that the $3$-way supermap in \eqref{eq_app:3waysol} admits a simpler realisation: it's a convex combination of single-qubit gates applied at the output of the channel. 
This motivates us to look for a similar realisation for the $4$-way supermap found in \eqref{eq_app:4waysol}. We look for a generalisation of the $3$-way realisation: a convex combination of single-qubit Pauli gates applied both before and after the channel action. That is, we look for a supermap represented by a positive operator of the following form
\begin{equation}\label{eq_app:Paulisupermap}
S_{\text{Pauli}} = \sum_{\vec{\gamma}, \vec{\delta}} p( \vec{\gamma}, \vec{\delta}) \, \,  S_{\vec{\gamma}, \vec{\delta}}  \,  
\end{equation}
where $p( \vec{\gamma}, \vec{\delta}) $ is a probability distribution and
\begin{equation}
 S_{\vec{\gamma}, \vec{\delta}} = \sigma_{\gamma_1} \otimes \sigma_{\gamma_2} \otimes \sigma_{\delta_1} \otimes \sigma_{\delta_2}  \otimes  \mathbbm{1}_H ( \ketbra{\Phi^+}{\Phi^+}_{HK} )  \, 
\end{equation}
is the positive operator in the form \eqref{eq_app:superchoi} corresponding to the map that applies the Pauli gates $\sigma_{\gamma_i/ \delta_i } (\rho) = \sigma_{\gamma_i/ \delta_i } \rho \sigma_{\gamma_i/ \delta_i } $ before and after the channel, with $\gamma_i, \delta_i = X,Y,Z$.

Notice that any supermap of the form \eqref{eq_app:Paulisupermap} satisfies conditions \eqref{eq_app:rescaling} by construction. Indeed, the action of Pauli gates on strings of Pauli operators can only amount to a change of sign. Given the supermap defined by the rescaling parameters in \eqref{eq_app:4waysol}, we want to determine if it can be written as \eqref{eq_app:Paulisupermap}, for a given choice of $p( \vec{\gamma}, \vec{\delta})$. Interestingly, the problem can be cast as a linear program \cite{boyd2004convex}.
For any choice of $\vec{\alpha}, \vec{\beta}$, define the rescaling factor of the target supermap as $x_{\vec{\alpha}, \vec{\beta}}^{\text{target}}$. Similarly, denote the corresponding rescaling factor for the Pauli supermap $S_{\vec{\gamma}, \vec{\delta}}$ as  $x_{\vec{\alpha}, \vec{\beta}}^{\vec{\gamma}, \vec{\delta}}$.
Then, the problem of representing the target supermap as a convex combination of Pauli supermap \eqref{eq_app:Paulisupermap} corresponds to finding a distribution $p( \vec{\gamma}, \vec{\delta})$ satisfying
\begin{align}
x_{\vec{\alpha}, \vec{\beta}}^{\text{target}}  = &  \sum_{\vec{\gamma}, \vec{\delta}}  p( \vec{\gamma}, \vec{\delta}) \, \,   x_{\vec{\alpha}, \vec{\beta}}^{\vec{\gamma}, \vec{\delta}} \quad \, \, \forall \vec{\alpha}, \vec{\beta} \nonumber \\
\text{s.t.} \quad &  p( \vec{\gamma}, \vec{\delta}) \geq 0 \quad \forall \, \, \vec{\gamma}, \vec{\delta} \nonumber \\
&  \sum_{\vec{\gamma}, \vec{\delta}}  p( \vec{\gamma}, \vec{\delta}) = 1\, .
\end{align} 
This is an instance of a linear program that can be solved with standard convex optimisation packages such as CVXPY. For a clearer interpretation of the solution,  we solved the above problem setting where the target rescaling parameters $x_{\vec{\alpha}, \vec{\beta}}^{\vec{\gamma}, \vec{\delta}}$ are those coming from the two extremal solutions in \eqref{eq_app:4waysol}, corresponding to $\lambda = 0,1$. For $\lambda = 0$ we get a representation with the following distribution
\begin{align}
    p(\gamma_1 \mathbbm{1} , \mathbbm{1} \delta_2 ) & = \frac{1}{12}  \quad \text{for} \, \, \gamma_1,\delta_2 = X,Y,Z \\
    p(\mathbbm{1} \mathbbm{1} , \mathbbm{1} \mathbbm{1}) & = \frac{1}{4} \nonumber \, .
\end{align}
Similarly, for $\lambda = 1$ we get
\begin{align}\label{eq_app:4wayPauli}
    p( \mathbbm{1} \gamma_2 , \delta_1 \mathbbm{1}  ) & = \frac{1}{12}  \quad \text{for} \, \, \gamma_2,\delta_1 = X,Y,Z \\
    p(\mathbbm{1} \mathbbm{1} , \mathbbm{1} \mathbbm{1}) & = \frac{1}{4} \nonumber \, .
\end{align}
Now it is evident that \eqref{eq_app:4wayPauli} corresponds to the twirling strategy described in \eqref{eq:twirling_4way}.

\section{Parameters of the gates used for numerical simulations}\label{app:parameters}

We used the gates from the 4-way reflection strategy 
with parameters $\theta_z=0.6, \delta_x=\delta_y=0.05$.
The single-site gates were parametrized as 
\be
W(x, y, z) = e^{i(\alpha\sigma_X + \beta\sigma_Y + \gamma\sigma_Z)},
\label{eq:param}
\ee
where $\alpha,\beta,\gamma$ are real parameters.
These values are reported in Tab.~\ref{tab:gate_params}.
\begin{table}[h!]
\centering
\caption{Numerical parameters for the single site $SU(2)$ gates from Eq.\eqref{eq:Ui} given by parametrization Eq.\eqref{eq:param}. Each gate is defined by the three parameters $(\alpha, \beta, \gamma)$. The first (second) two gates correspond to even (odd) layers of the time evolution.}
\label{tab:gate_params}
\begin{tabular}{@{}lcccc@{}}
\toprule
\textbf{Parameter} & \textbf{Gate $W_B$} & \textbf{Gate $W_A$} & \textbf{Gate $W_B'$} & \textbf{Gate $W_A'$} \\
\midrule
$\alpha$ & 1.64979 & 1.54383& 0.45310 & 1.53416 \\
$\beta$  & 0.48791 & 1.80539 & 1.11250  & 0.20499 \\
$\gamma$ & 0.20562  & 0.17212 & 1.60546  & 1.04460  \\
\bottomrule
\end{tabular}
\end{table}

The observable is a normalized linear combination of Pauli matrices:
\be
 O_1  = \frac{0.57 \sigma_X + 0.12 \sigma_Y + 0.84 \sigma_Z}{\|0.57 \sigma_X + 0.12 \sigma_Y + 0.84 \sigma_Z\|_2},
\ee
where the norm  is calculated using the trace of the squared operator: $\|A\|_2 = \sqrt{\text{Tr}(A^\dagger A)}$.

\section{Pauli diagonal noise benchmarking}\label{sec_app:paulinoise}

We exemplify the use of average-computation benchmarking to estimate noisy expectation values instead of ideal ones. We show that, under some specific noise model, the presence of hardware noise can be added to the calculation while preserving the efficiency. The considered error model follows closely the one from the recent IBM quantum experiments \cite{kim2023evidence,yoshioka2025krylov}. After each layer of two-qubit gates, the hardware errors are represented by a noise channel $\mathcal{E}_{\text{noise}}$ affecting all qubits. The channel is assumed to be of a Pauli diagonal form, namely
\begin{equation}\label{eq:Pauli_channel}
\mathcal{E}_{\text{noise}} = \sum_{\vec{\alpha}} p_{\vec{\alpha}} \sigma_{\vec{\alpha}} ( \cdot ) \sigma_{\vec{\alpha}}    \, ,
\end{equation}
where $p_{\vec{\alpha}}$ is a normalised probability distribution, $\vec{\alpha} = \alpha_0,\ldots, \alpha_{L-1}$ and $\alpha_i = \mathbbm{1},X,Y,Z$ for all $i$. Interestingly, for channels of this form, the calculation of the two-point correlation function for average computation simplifies greatly even in the noisy case. 
Expectation values of the output of the noisy circuits correspond to
\begin{equation}\label{eq:two-point_noisy}
    \langle \mathcal{O}_{\text{test}} \rangle = \Tr (  \mathcal{O}_{\text{test}} \, \, \mathcal{E}_{\text{noise}} U_{l_T} \ldots U_{l_{T-1}+1} \mathcal{E}_{\text{noise}} \ldots U_{l_1 + 1} \mathcal{E}_{\text{noise}} U_{l_1} \ldots U_1 (  \ket{\psi_0 }\bra{\psi_0}) ) \, ,
\end{equation}
where we denote $U_{l_i}$ as the last gate of the $i$-th brickword layer. For what follows, we assume that the noise channel does neither depend on the gate nor on the specific layer of the circuit. Hence, the noisy correlation functions for the average computation can be computed as
\begin{equation}\label{eq:two-point_noisy_avg}
    \langle \mathcal{O}_{\text{test}} \rangle = \Tr (  \mathcal{O}_{\text{test}} \, \, \mathcal{E}_{\text{noise}} \mathcal{E}_{l_T} \ldots \mathcal{E}_{l_{T-1}+1} \mathcal{E}_{\text{noise}} \ldots \mathcal{E}_{l_1 + 1} \mathcal{E}_{\text{noise}} \mathcal{E}_{l_1} \ldots \mathcal{E}_1 (  \ket{\psi_0 }\bra{\psi_0}) ) \, ,
\end{equation}
where $\mathcal{E}_i$ represents the space-time channel corresponding to the average computation for gate $U_i$.
We show now that the channel defined in \eqref{eq:Pauli_channel} satisfies useful properties that allow to simplify the above calculation even in the noisy case. In particular, it holds
\begin{equation}\label{eq_app:identityupwards}
\Tr_i ( \sigma_{\alpha}^{(i)}  \, \mathcal{E} (o_{\text{rest}} \otimes \mathbbm{1}_i) )  = \delta_{\alpha \mathbbm{1}} \, \Tr_i ( \mathcal{E} (o_{\text{rest}} \otimes \mathbbm{1}_i) )  
\end{equation}
and 
\begin{equation}\label{eq_app:identitydownwards}
\mathrm{Tr}_i ( \mathcal{E} (o_{\text{rest}} \otimes \sigma_{\beta}^{(i)} ) ) = \delta_{\beta \mathbbm{1}} \, \mathrm{Tr}_i ( \mathcal{E} (o_{\text{rest}}\otimes \mathbbm{1}_i) ) \, .
\end{equation}
Graphically, they can be represented as follows

\begin{equation}
\begin{tikzpicture}[baseline=(current  bounding  box.center), scale=0.6]

 \foreach \i in {0,...,9}{
 \draw[very thick] (\i,-.5) -- (\i,.5);
 }
 \Noise{0-.25}{9+.25}{0}
\MYcircle{5}{-.5}
\end{tikzpicture}
=
\begin{tikzpicture}[baseline=(current  bounding  box.center), scale=0.6]
\foreach \i in {0,...,9}{
 \draw[very thick] (\i,-.5) -- (\i,.5);
 }
 \Noise{0-.25}{4+.25}{0}
 \Noise{6-.25}{9+.25}{0}
\MYcircle{5}{-.5}
\end{tikzpicture}
\end{equation}
and
\begin{equation}
  \begin{tikzpicture}[baseline=(current  bounding  box.center), scale=0.6]

 \foreach \i in {0,...,9}{
 \draw[very thick] (\i,-.5) -- (\i,.5);
 }
 \Noise{0-.25}{9+.25}{0}
\MYcircle{5}{.5}
\end{tikzpicture}
=
\begin{tikzpicture}[baseline=(current  bounding  box.center), scale=0.6]
\foreach \i in {0,...,9}{
 \draw[very thick] (\i,-.5) -- (\i,.5);
 }
 \Noise{0-.25}{4+.25}{0}
 \Noise{6-.25}{9+.25}{0}
\MYcircle{5}{.5}
\end{tikzpicture}.
\end{equation}

One can interpret them as the fact that Pauli diagonal channels allow for the propagation of the identity on every site, both backwards and forward in time. This feature is enough to allow performing all the simplification procedures used to compute correlations for $4$-way and $3$-way channels, as outlined in the End Matter. 
Let us start with the case of two-body correlation functions and compute the noisy equivalent of \eqref{eq:quench5}. First, if we include the noise channels in \eqref{eq:quench2}, we obtain
\begin{align}\label{eq:quench2}
&\langle O^{(i,i+2T+1)} \rangle_\text{avg} = \begin{tikzpicture}[baseline=(current  bounding  box.center), scale=0.6]
\foreach \i in {0,...,1}{
\Wgategreen{-2}{0+3*\i}
\Wgategreen{0}{0+3*\i}
\Wgategreen{2}{0+3*\i}
\Wgategreen{4}{0+3*\i}
\Wgategreen{6}{0+3*\i}
\Wgategreen{-1}{2-.5+3*\i}
\Wgategreen{1}{2-.5+3*\i}
\Wgategreen{3}{2-.5+3*\i}
\Wgategreen{5}{2-.5+3*\i}
\Wgategreen{7}{2-.5+3*\i}
}
\foreach \i in {2.5,4.5,6.5,8.5,10.5}
{\Bell{-4+\i}{-.5}}
\Noise{-2.75}{7.75}{.75}
\Noise{-2.75}{7.75}{3-.25-.5}
\Noise{-2.75}{7.75}{5-1-.25}
\Noise{-2.75}{7.75}{7-1.5-.25}
 \foreach \i in {-1,...,8}{
 \draw[very thick] (\i-0.5,5.5) -- (\i-0.5,5.75);
 }
  \foreach \i in {-1,...,8}{
 \draw[thick, fill=white] (\i-0.5,5.75) circle (0.1cm); 
}
\MYcircleB{7.5}{5.75}
\MYcircleB{-1.5}{5.75}
\Text[x=-1.5,y=6.25]{$a$}
\Text[x=7.5,y=6.25]{$b$}
\end{tikzpicture}.
\end{align}
Combining trace preservation and the property \eqref{eq_app:identitydownwards}, one obtains a modified backwards lightcone diagram
\begin{align}
\langle O^{(i,i+2T+1)} \rangle_\text{avg}=
\begin{tikzpicture}[baseline=(current  bounding  box.center), scale=0.6]
\Wgategreen{-4}{0}\Wgategreen{-2}{0}
\Wgategreen{0}{0}\Wgategreen{2}{0}\Wgategreen{4}{0}\Wgategreen{6}{0}\Wgategreen{8}{0}\Wgategreen{10}{0}
\Wgategreen{-1}{3+1.5}\Wgategreen{7}{3+1.5}
\Wgategreen{-2}{2+1}\Wgategreen{0}{2+1}\Wgategreen{6}{2+1}\Wgategreen{8}{2+1}
\Wgategreen{-3}{1+.5}\Wgategreen{-1}{1+.5}\Wgategreen{1}{1+.5}\Wgategreen{5}{1+.5}\Wgategreen{7}{1+.5}\Wgategreen{9}{1+.5}
\foreach \i in {-1.5,0.5,2.5,4.5,6.5,8.5,10.5,12.5,14.5}
{
\Bell{-3.5+\i-.5}{-.5}
}
\Noise{-3.5}{1.5}{.75}
\Noise{4.5}{9.5}{.75}
\Noise{-3.5+1}{1.5-1}{.75+1.5}
\Noise{4.5+1}{9.5-1}{.75+1.5}
\Noise{-3.5+2}{1.5-2}{.75+3.}
\Noise{4.5+2}{9.5-2}{.75+3}
\Noise{-3.5+2}{1.5-3}{.75+4.5}
\Noise{4.5+3}{9.5-2}{.75+4.5}
 \foreach \i in {0,...,3}{
\MYcircle{\i+4-.5}{0.5+1.5*\i}
\MYcircle{\i-.5}{3.5-1.5*\i+1.5}
}
 \foreach \i in {1,...,3}{
\MYcircle{\i-.5+8}{3.5-1.5*\i+1.5}
\MYcircle{\i-4-.5-1}{0.5+1.5*\i-1.5}
}
 \MYcircle{-5.5}{-.5}
 \MYcircle{11.5}{-.5}
  \foreach \i in {7.5,-1.5}{
  \draw[very thick] (\i,5.5) -- (\i,5.75); 
}
\MYcircleB{7.5}{5.75}
\MYcircleB{-1.5}{5.75}\Text[x=-1.5,y=6.25]{$a$}
\Text[x=7.5,y=6.25]{$b$}
\end{tikzpicture}.
\label{eq:quench4new}
\end{align}
Lastly, iterating right and left unitality, combined with \eqref{eq_app:identityupwards} leads to
\begin{align}
\langle O^{(i,i+2T+1)} \rangle_\text{avg}=
\begin{tikzpicture}[baseline=(current  bounding  box.center), scale=0.6]
\Wgategreen{2}{0}\Wgategreen{4}{0}
\Wgategreen{-1}{3+1.5}\Wgategreen{7}{3+1.5}
\Wgategreen{0}{2+1}\Wgategreen{6}{2+1}
\Wgategreen{1}{1+.5}\Wgategreen{5}{1+.5}
\Bell{-3.5+6.5-.5}{-.5}
\Noise{1.5}{1.5}{.75}
\Noise{4.5}{4.5}{.75}
\Noise{.5}{0.5}{.75+1.5}
\Noise{5.5}{5.5}{.75+1.5}
\Noise{-0.5}{-0.5}{.75+3.}
\Noise{4.5+2}{6.5}{.75+3}
\Noise{-1.5}{-1.5}{.75+4.5}
\Noise{4.5+3}{9.5-2}{.75+4.5}
 \foreach \i in {0,...,3}{
\MYcircle{\i+4-.5}{0.5+1.5*\i}
\MYcircle{\i-.5}{3.5-1.5*\i+1.5}
\MYcircle{\i+4+0.5}{-0.5+1.5*\i}
\MYcircle{\i-1.5}{3.5-1.5*\i+0.5}
}
  \foreach \i in {7.5,-1.5}{
  \draw[very thick] (\i,5.5) -- (\i,5.75); 
}
\MYcircleB{7.5}{5.75}
\MYcircleB{-1.5}{5.75}\Text[x=-1.5,y=6.25]{$a$}
\Text[x=7.5,y=6.25]{$b$}
\end{tikzpicture}.
\label{eq:quench5new}
\end{align}
One can see the above equation as the equivalent of \eqref{eq:quench5} for noisy two-body correlation functions. Written as an explicit equation, it reads
\begin{align}
  \langle O^{(i,i+2T+1)} \rangle_\text{avg}=\bra{a b}  \prod_{k=1}^T  (E{_\text{noise}})_k ( \mathcal{M}_-)_k \otimes (E{_\text{noise}})_k (\mathcal{M}_+)_k\ket{\rho_{\text{Bell}}},
\end{align}
where $(E{_\text{noise}})k$ is the reduced channel acting on the output particle of the transfer matrix at layer $k$. Notice that we defined the reduced channel acting on particle $i$ as
\begin{equation}
    E_{\text{red}}^{(i)} ( \cdot ) = \Tr_{I / i} \mathcal{E}_{\text{noise}}( \mathbbm{1}_0 \otimes \ldots \mathbbm{1}_{i-1} \otimes \cdot \otimes  \mathbbm{1}_{i+1} \otimes \ldots  \mathbbm{1}_{L-1}  ) .
\end{equation}
Proceeding similarly, the noisy version of the single-particle expectation values \eqref{eq:quench6} becomes
\begin{equation}
\langle O^{(T)} \rangle_\text{avg}=
\begin{tikzpicture}[baseline=(current  bounding  box.center), scale=0.6]
\Wgategreen{3}{4.25}
\Wgategreen{2}{2.75}
\Wgategreen{1}{1.25}
\Wgategreen{0}{-0.25}
\Noise{.5}{.5}{.5}
\Noise{1.5}{.5+1}{.5+1+.5}
\Noise{.5+2}{.5+2}{.5+3}
\Noise{.5+3}{.5+3}{.5+4.5}
\foreach \i in {0,...,3}{
\MYcircle{\i-.5}{\i*1.5+.25}
\MYcircle{\i+2-.5+1-2}{0.75+\i*1.5-1.5}
}
\draw[very thick] (3.5,5.25) -- (3.5,5.5);
\MYcircleB{3.5}{5.5}
\Text[x=-0.5,y=-1.]{$+$}
\Text[x=3.5,y=6.0]{$O$}
\end{tikzpicture}.
\label{eq:quench6}
\end{equation}
which corresponds to the following equation
\begin{equation}
    \langle O^{(T)} \rangle_\text{avg} = \text{Tr} \; ( O_{T} \sigma_X^{(0)}(T) ) = \bra{O}\prod_{k=1}^T(E_{\text{noise}})_k(\mathcal{M}_+)_k\ket{\sigma_X}.
\end{equation}

Lastly, the noisy version of the nearest-neighbor correlation functions \eqref{eq:quench7} undergoes a similar modification:  the reduced channel is interleaved between layers of transfer matrices. For the sake of brevity, we don't report its explicit expression.  

\end{document}